\documentclass{article}

\usepackage{arxiv}

\usepackage[utf8]{inputenc} 
\usepackage[T1]{fontenc}    
\usepackage{hyperref}       
\usepackage{url}            
\usepackage{booktabs}       
\usepackage{amsfonts}       
\usepackage{nicefrac}       
\usepackage{microtype}      
\usepackage{lipsum}
\usepackage{graphicx}
\graphicspath{ {./images/} }

\usepackage{multirow}
\usepackage{subfigure}
\usepackage{hhline}
\usepackage{amsmath}

\usepackage{amssymb}

\hypersetup{hidelinks}
\usepackage{algorithm}
\usepackage{algpseudocode}

\title{Forecasting formation of a Tropical Cyclone Using Reanalysis Data}

\author{
 Sandeep Kumar \\
   Department of Mathematics, Shaheed Bhagat Singh College,\\ University of Delhi.\\ \& \\
  Department of Computer Science, IIIT Delhi\\ 
  New Delhi, India. \\
  \texttt{ sandeep\_kumar@sbs.du.ac.in, sandeepk@iiitd.ac.in} 
   \And
    Koushik Biswas\\
  Department of Computer Science \\
  IIIT Delhi\\
  New Delhi, India, 110020 \\
  \texttt{koushikb@iiitd.ac.in} \\
   \And
 Ashish Kumar Pandey \\
  Department of Mathematics\\
  IIIT Delhi\\
  New Delhi, India, 110020 \\
  \texttt{ashish.pandey@iiitd.ac.in} \\
}

\date{}

\begin{document}
\maketitle
\begin{abstract}
The tropical cyclone formation process is one of the most complex natural phenomena which is governed by various atmospheric, oceanographic, and geographic factors that varies with time and space. Despite several years of research, accurately predicting tropical cyclone formation remains a challenging task. While the existing numerical models have inherent limitations, the machine learning models fail to capture the spatial and temporal dimensions of the causal factors behind TC formation. In this study, a deep learning model has been proposed that can forecast the formation of a tropical cyclone with a lead time of up to 60 hours with high accuracy. The model uses the high-resolution reanalysis data ERA5 (ECMWF reanalysis $5^{th}$ generation), and best track data IBTrACS (International Best Track Archive for Climate Stewardship) to forecast tropical cyclone formation in six ocean basins of the world. For 60 hours lead time the models achieve an accuracy in the range of $86.9\% - 92.9\%$ across the six ocean basins. The model takes about 5-15 minutes of training time depending on the ocean basin, and the amount of data used and can predict within seconds, thereby making it suitable for real-life usage.
\end{abstract}


\section{Introduction}

The formation of any natural disaster is a complicated phenomenon that involves multiple causal factors which have temporal, spatial, and altitudinal dimensions. Understanding the evolution process of a natural disaster and modeling it, is always a challenging task. One such natural disaster is tropical cyclone(TC) (also known as hurricanes or typhoons) which occurs frequently in the tropical and subtropical waters of the world. Near the equator, the warm air rises over the surface of the sea and creates a low-pressure system (LPS), also known as a tropical depression. This causes the air around the LPS to move towards it, which further gets warmed and rises above.  The rising moist air cools down and forms the cloud. The process of cloud formation and wind rotation intensifies with the help of favorable conditions like- sea surface temperature greater than $26^{o}C$ \cite{pal1948for}, low vertical wind shear,  high relative humidity, and atmospheric instability~\cite{pal1948for,elsberry1995tropical,gray1968global}. The difference of temperature between the warm core with rising moist air and the adjoining cool environment leads to rapidly rising buoyant air. All this makes the TC development a complicated process that depends on oceanographic, atmospheric, and geographic factors. Moreover, out of these LPSs, only a small number developed in a full-fledged TC under the above favorable conditions \cite{emanuel1989finite}.  As the theory behind the development process of a TC is still not settled, predicting TC formation is a challenging problem. TCs bring with themselves heavy rainfall, thunderstorm, and flash floods in the coastal areas, thereby causing huge ecological, infrastructural and human loss.  Also with climate change, the frequency, intensity, and associated hazards of TCs are going to increase \cite{tclkmm2019tcacc}. All this makes, the development of a model that can forecast the formation of a TC well advance in time,  important from a disaster mitigation point of view. This will provide the disaster managers adequate time to take preventive measures. In this work, a deep learning model has been proposed to successfully forecast TC formation with a lead time of up to 60 hours (h) using as less as 12h of preceding data for the six ocean basins, North Atlantic (NA), North Indian (NI), South Indian (SI), West Pacific (WP), South Pacific (SP), and East Pacific (EP) of the world. 

There are mainly two approaches for detecting TCs, one is model driven approach based on equations governing the physical phenomenon of TC development (including numerical simulations) and the other is data driven approach that utilizes historical data relating to TCs (including machine learning methods). The earliest conventional way to detect a TC is early-stage Dvorak analysis (EDA) (an extended Dvorak technique \cite{dvorak1975}) which utilizes satellite cloud images, however this technique includes subjective interpretation of parameters and hence not sufficiently scientific \cite{dvorak1984}. EDA is used by the National Hurricane Center (NHC), Central Pacific Hurricane Center (CPHC), and the Japan Meteorological Agency (JMA) to forecast typhoon initiation, up to 48 hours before its formation, with an accuracy of up to $57\%$ \cite{tcfg2013jc}.  In \cite{tcgg2017myn}, authors shows that the global ensembles models [European Centre for Medium-Range Weather Forecasts (ECMWF), Japan Meteorological Agency (JMA), National Centers for Environmental Prediction (NCEP), and Met Office in the United Kingdom (UKMO)] along with EDA can be used to improve the accuracy up to $79\%$.
In \cite{halp2013eval}, authors evaluated the performance of five global NWP systems [Global Forecast System (GFS), ECMWF, Canada's Global Environment Multiscale Model (CMC), UKMO, and Navy Operational Global Atmospheric Prediction System (NOGAPS)] for TC forecast in the North Atlantic (NA) ocean for the period 2004-2011, and shows the best hit rate of $44\%$ can be achieved. 
Over the years the accuracy of numerical models is improved based on better initialization and improved computing power. But still, these models are not suitable for long lead time forecasts as the numerical methods are prone to error accumulation over iterations. In \cite{oeptc2009as,henn2003for} authors have used statistical method, Linear Discriminant Analysis (LDA) to predict 24h probability of TC formation from derived large scale environmental parameters. But the factors that lead to a TC formation have a non-linear complex relationship that makes these linear methods unsuitable for the TC formation prediction task. Recently, machine learning methods and deep learning methods have been successfully applied to answer the TC formation forecast problem, which we will discuss more in the next section. 

The rest of the paper is organized as follows: in Section~\ref{sec2} related work is described, Section~\ref{sec3} presents the data used, Section~\ref{sec4} describe the proposed deep learning model, Section~\ref{sec5} presents the findings of this work, and finally in Section~\ref{sec6} we conclude with a summary and future directions. 

\section{Related Work} \label{sec2}

Various machine learning models have been successfully applied to a TC formation forecast problem. In \cite{ddntd2015wzfm} Decision trees (DTs) are used to detect developing and non-developing tropical disturbances in the North Pacific Ocean for the months, June to September of 2004-2013, using five derived parameters from Navy Operational Global Atmospheric Prediction System (NOGAPS). They reported accuracy of $84.6\%$ for a lead time of 24h. They differentiate between developing and non-developing disturbance based on relative vorticity. In \cite{rs11101195}, DTs, random forest (RF), and support vector machine (SVM) are used to detect the formation of TC in the western North Pacific Ocean for the period 2005-2009 using eight derived predictors from WindSat satellite data. They classify a tropical depression as  TC when the maximum sustained wind speed (MSWS) reaches $13m/s$ (or 25 knots) and the satellite image available with at least $60\%$ coverage in a circle with $4^\circ$ radius around the center of the tropical disturbance. In \cite{ptcgmc2019tzwy}, the authors use 13 predictors derived from mesoscale convective system (MCS) data and ERA-Interim dataset to predict TC formation using the following machine learning tools - Logistic Regression (LR), Naïve Bayes (NB), DT, K-Nearest neighbors (KNN), Multilayer perceptron (MLP), Quadratic Discriminant Analysis (QDA), SVM, AdaBoost (ADA), and RF. The authors reported the accuracy in terms of F1-score, precision, and recall for lead times 6h, 12h, 24h, and 48h for global (consisting of all ocean basins), NA, and west north pacific (WNP) ocean basins.

Various deep learning  studies have successfully captured the spatial and temporal dimensions of causal factors to answer the prediction problems related to TC's track, intensity \cite{abjags2018phturnn,gspymjzsz2018anmpttlstm,kumar2021tptulstm,buobhlr2019,mmrrmgieepti2020, kumar2021track} and its landfall's characteristics \cite{kumar2021imd,kumar2021reanalysis, kumar2022will}. Recently, a few deep learning studies have been proposed that forecast TC formation. In \cite{matsuoka2018deep}, CNN has been used to detect the TC and its precursors in the six ocean basins of the world using 30 years of simulated outgoing longwave radiation (OLR) data generated through a cloud-resolving global non-hydrostatic atmospheric model. The TC and its precursors are categorized as one class and identified based on TC tracking algorithm \cite{sugi2002influence,yam2017res} which takes temperature, horizontal components of wind, and sea level pressure (SLP) as inputs. The study restricted it to the limited range of latitudes $30^\circ$S - $30^\circ$N. In \cite{dldib2020sssgm}, the authors have presented a deep learning model to detect an ongoing TC with the help of satellite data of eight TCs in NI ocean basin. In \cite{chen2019hybrid} a hybrid CNN-LSTM model is used to predict if an ongoing TC will be intensified to the level of a typhoon (wind speed greater than 64 knots) with a lead time of 24h, using preceding data of 6h, 12h, 18h or 24h. The International Best Track Archive for Climate Stewardship (IBTrACS) data and ERA-Interim datasets are used for three ocean basins WP, EP, and NA. Thus we see that very few deep learning studies exists, and each has their own criterion of detecting TC formation. 

\section{Data} \label{sec3}

\begin{figure*}[!h]
    \centering
    \caption{Location of TC and non TC formation.}
    \includegraphics[width = 17cm, height = 6cm]{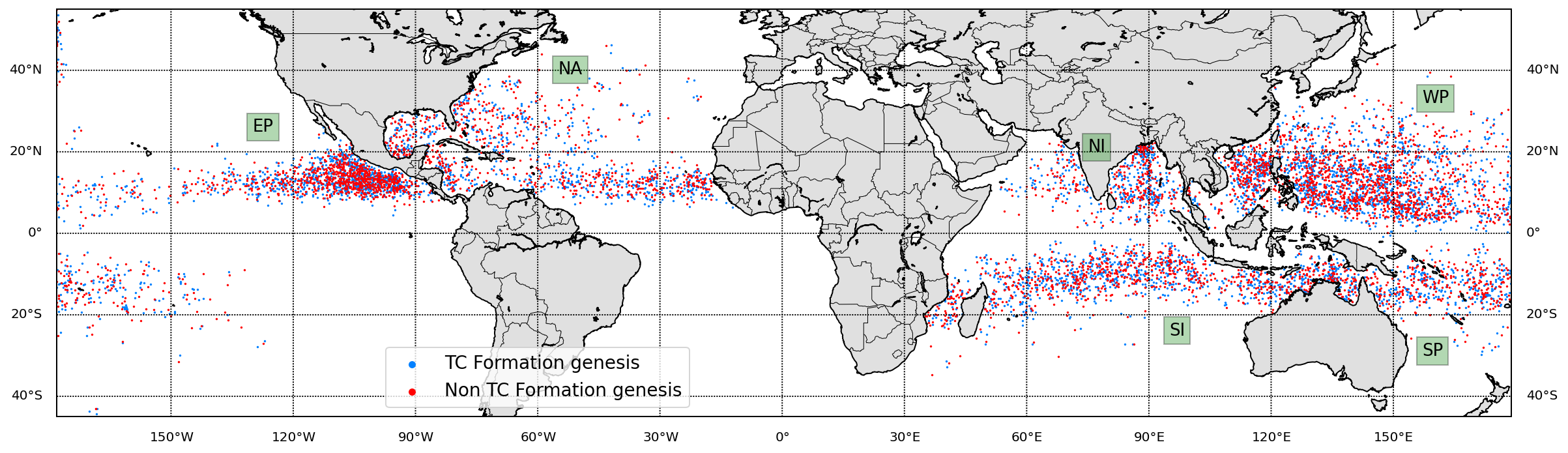}
    \label{data1}
\end{figure*}

\begin{figure}[!h]
    \centering
    \caption{Pictorial representation of reanalysis data.}
    \includegraphics[width = 7cm, height = 8cm]{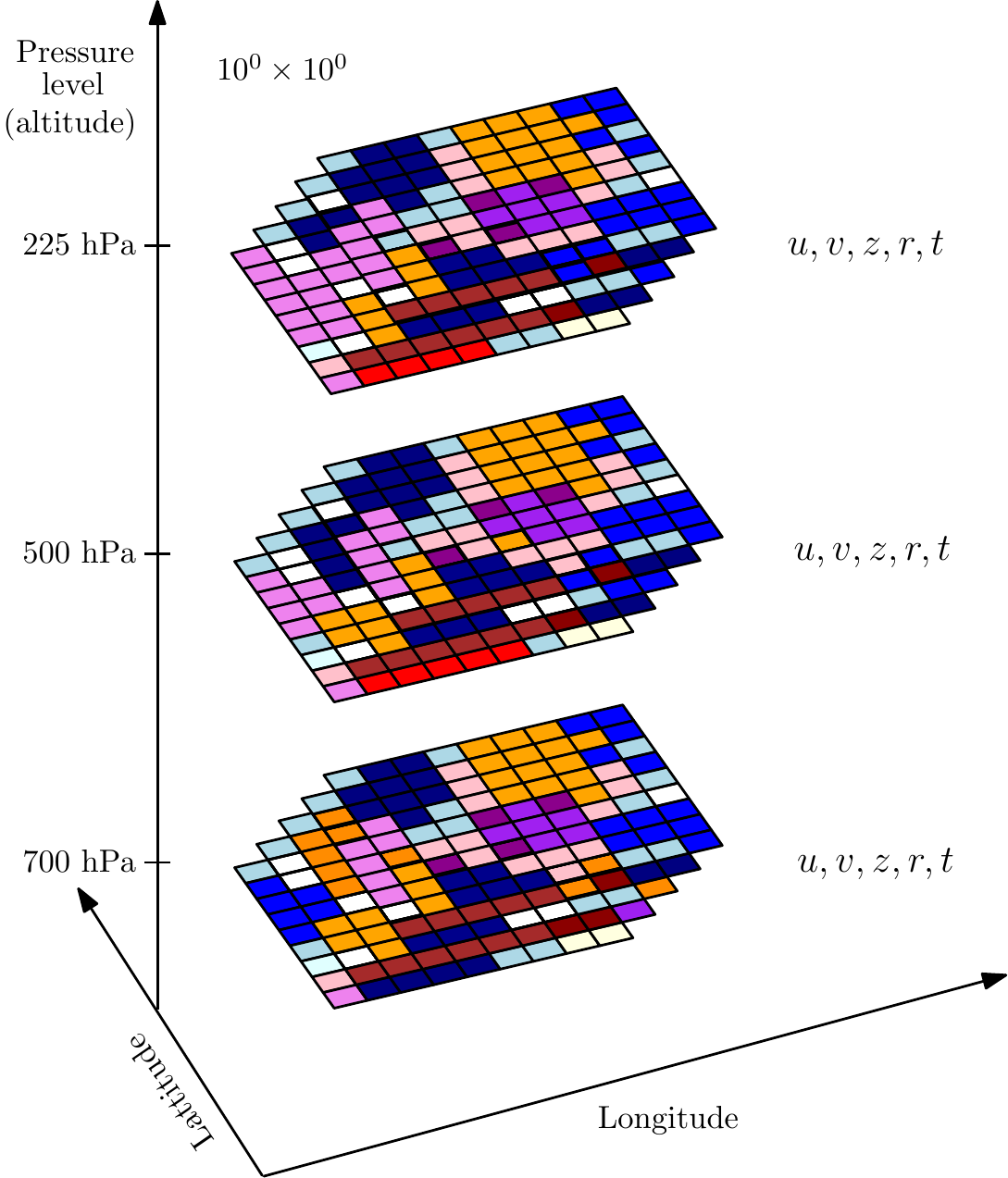}
    \label{data3}
\end{figure}

Inspired by the successful usage of reanalysis dataset in the recent works \cite{chen2019hybrid,kumar2021reanalysis,boussioux2020hurricane,gsygk2020tctufl, kumar2022will} to answer TC related track, intensity, and landfall's characteristics problems, we have used ERA5 \cite{era5cite} high-resolution reanalysis dataset provided by ECMWF\footnote{\url{https://cds.climate.copernicus.eu/}}, a high-resolution data that provides hourly weather and climate data for the whole globe. In \cite{hodges2017well}, the authors show how well a TC is represented in the reanalysis dataset. The formation process of a TC is determined by large-scale atmospheric factors at various altitudes around the center of a LPS. For this study we have extracted wind fields $u$, $v$, geopotential $z$, relative humidity $r$, and temperature $t$ at three pressure levels (altitudes) 225hPa, 500hPa, and 700hPa. These variables largely determine the development process of a TC as follows: $u$ and $v$ fields represent the east-west and north-south movement of air along with its speed, $z$ represents the gravitational potential energy relative to sea level, $r$ represents the water vapor pressure, and $t$ represents the atmospheric temperature. As the atmospheric causal factors behind TC formation may have horizontal extends up to 1000 kilometers (km) these variables are extracted for a spatial region $5^\circ \times 5^\circ$ with a resolution of $0.25^\circ$ which resulted in a grid of $41\times 41$. As one degree is around 110 km near the equator and decreases as we move pole-wards, this resulted in a spatial extend of around 1000 km with a resolution of around 25 km. A graphical representation of used reanalysis data is shown in Fig~\ref{data3}.

\begin{algorithm}
\caption{Generating negative classes (non TC) in a ocean basin, say \textit{OB}.}
\label{alg:algorithm}
\textbf{Input}: Set $T$ and $L$ of time points and location (lat, lon) of all TC formation in ocean basin \textit{OB}. \\
\textbf{Parameter}: \textit{count} (No. of positive class in \textit{OB}), A Kernel Density Function say \textit{LocGen} fitted on the set $L$.\\
\textbf{Output}: Set $T1$ and $L1$ of time points and location \textit{(lat, lon)} of all non TC formation (negative classes) in ocean basin \textit{OB}.
\begin{algorithmic}[1] 
\State Let $count1=0$, $T1 = \{\},$, $L1 = \{\}$.
\While{$count1<count$}
    \State Generate a random time $t1$ (between 01/01/1980 and 01/09/2021) and random location say \textit{(lat1,lon1)}  through \textit{LocGen}.
    \If{\textit{(lat1,lon1)} lies over land}
        \State continue
    \EndIf
    \If{$Abs(t-t1) >$  5days $\forall t \in T$}
        \State Add $t1$ to $T1$ and \textit{(lat1,lon1)} to $L1$, $count1++$.
    \Else
        \If{$Abs(lat-lat1) > 5^\circ$ and  $Abs(lon-lon1) > 5^\circ$ $\forall \, (lat, lon) \in L$.}
            \State Add $t1$ to $T1$ and \textit{(lat1, lon1)} to $L1$, $count1++$.
        \EndIf
    \EndIf
\EndWhile
\State \textbf{return} $T1$ and $L1$.
\end{algorithmic}
\end{algorithm}

The IBTrACS dataset \cite{ibtrackdata} maintained by National Oceanic and Atmospheric Administration\footnote{\url{https://www.ncdc.noaa.gov/ibtracs/}} keeps three hourly global records of all TCs in the form of its time, track (latitude and longitude), intensity, and many more other variables from the very initiation of a TC when it was first detected as a tropical depression or LPS. As the definition of TC genesis time is ambiguous \cite{tsd2014mhk} we take the time when a TC is recorded first as LPS in IBTrACS as  TC formation time. We extracted the record of TC formation time and corresponding location (latitude, longitude) for all TCs for the earlier mentioned six ocean basins of the world from 1980 to September 2021. All these TCs form the positive class in our classification problem. The total number of positive cases are 653 (NA), 360 (NI), 832 (SI), 1431 (WP), 509 (SP), 983 (EP). To generate the negative classes (non-TC formation data), for a particular ocean basin we followed the Algorithm~\ref{alg:algorithm}. This way we have equal number of positive and negative classes in our dataset. A negative class sample represents a time $t$ and location \textit{loc} such that there is no existing TC formation in a time window of 5 days and if $t$ lies within a window of 5 days, then there is no existing TC formation in a spatial window of $5^\circ$. This way we have selected all TC formation samples and non-TC formation samples. Next, for each sample, we downloaded the above described reanalysis data for grid size $41\times 41$ centered at the location of each sample,  for time points $t-6k$, $12 \geq k \geq 4 $, where $t$ is the time of TC formation or randomly selected time of a non-TC formation sample. Thus the reanalysis dataset is extracted for 9 time points from $t-72$h to $t-24$h at an interval of 6 hours. The genesis location of all samples (TC formation and non-TC formation) are shown in Fig~\ref{data1} for all six ocean basins.

\subsection{Training Dataset Preparation}

For a TC, 9 data points are available at an interval of 6h as described above. Suppose we want to use $T$ number of data points ((T-1)*6 hours of data) to predict the formation of a TC. For this, we generate $10-T$ training data points, where a single training point is a sequence of $T$ vectors of the form:
\begin{align*}
(&\operatorname{u225}(t),  \operatorname{v225} (t) \operatorname{z225} (t),  \operatorname{r225}(t), \operatorname{t225} (t), \\ & \operatorname{u500}(t),  \operatorname{v500} (t) \operatorname{z500} (t),  \operatorname{r500}(t), \operatorname{t500} (t), \\ & \operatorname{u700}(t),  \operatorname{v700} (t) \operatorname{7500} (t),  \operatorname{r700}(t), \operatorname{t700} (t))
\end{align*}
where $k\leq t\leq T+k-1$ and $k$ varies from $1$ to $10-T$. For each such training point, the target variable is 1 (positive class) or 0 (negative class).  One must note that the above process forms $10-T$ training points at leads hours $6k$ where $4\leq k \leq 9+4 -T$. All such training points for all the TCs form the training dataset. 

\section{Model and its Implementation} \label{sec4}

\begin{figure}[!h]
    \centering
        \caption{Model description for $T = 3$.}
    \includegraphics[width = 8cm, height = 17cm]{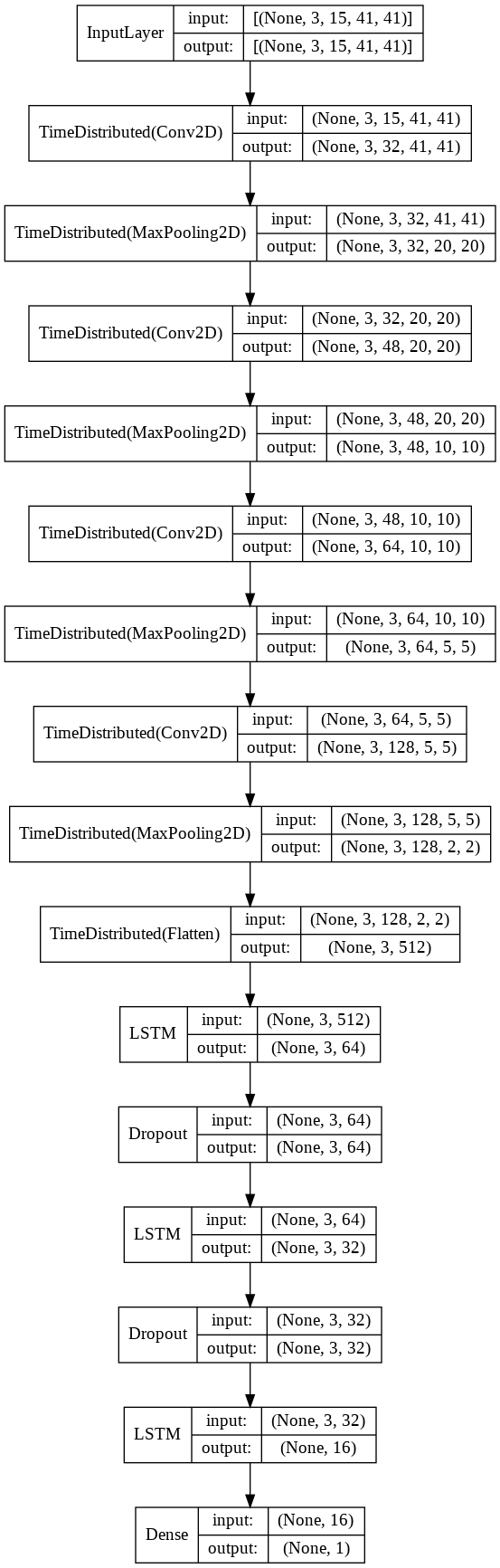}
    \label{model}
\end{figure}

As our dataset has both spatial and temporal dimensions, the model utilizes a combination of CNN \cite{6795724,kasihg2012icdcnn,7785132} and LSTM \cite{lstm0,lstm1,lstm2,lstm3}  networks to effectively capture the causal factors behind a TC formation.  The input training dataset is of the dimension $(T, 15, 41, 41)$, where $T$ stands for the length of sequential data points (of $6*(T-1)$h), 15 stands for the number of channels (corresponding to $u$, $v$, $z$, $r$, and $t$ fields at three pressure levels), and $(41, 41)$ is the shape of the grid centered at TC formation location. The model consists of four alternating convolution and max-pooling layers that generate sequential features of length $T$ using $\operatorname{TimeDistributed}$ layer of Keras \cite{chollet2015keras}, which are further fed into a stacked LSTM consisting of three LSTM layers. To avoid over-fitting dropout 0.15 is used between two successive LSTM layers. The model and input-output size of each layer is shown in Fig~\ref{model} for $T=3$. This resulted in a lightweight model with just 2,83,073  trainable parameters.

\subsection{Training and Implementation}

We experimented with various configurations of above model by varying number of layers and nodes in it, activation functions, and learning rates. The configuration which works well across all ocean basins is reported. The activation function $\operatorname{ReLU}(x) = \max(0, x)$ \cite{relu} is used in all layers except the last layer which uses $\operatorname{Sigmoid}(x) = \frac{1}{1+\exp(-x)}$ \cite{hjmsf1995inf} activation function. The input variables are scaled in the range $(-1, 1)$ for faster and stable training using $\operatorname{MinMaxScaler}$ of Scikit learn library \cite{scikit-learn}. The model uses optimizer $\operatorname{Adam}$ \cite{kdj2014adam}, default learning rate $0.001$, binary cross-entropy loss function, 32 batch size, and 30 epochs. The model is implemented in  $\operatorname{Keras}$ API developed over low-level language $\operatorname{TensorFlow}$ \cite{tensorflow2015-whitepaper} on Nvidia Tesla V100 GPU platform with 16 GB RAM, that takes around 5-15  minutes for 30 epochs depending on ocean basin and $T$. 

\subsection{Evaluation Metrics}

As reported in \cite{ptcgmc2019tzwy}, we have evaluated the performance of proposed model in terms of metrics - Precision, Recall, Accuracy and F1-score (F1) which are defined as:

$$Precision (P) = \frac{TP}{TP+FP}, \quad Recall (R) = \frac{TP}{TP+FN}$$
$$Accuracy = \frac{TP+TN}{TP+TN+FP+FN}, \quad F1 = \frac{2PR}{P+R}$$

\begin{table}[!h]
\begin{center}
\caption{Truth Table}
\begin{tabular}{p{2.5cm} p{2.5cm} p{1.5cm}}
\toprule%

&  \multicolumn{2}{c}{Predicted} \\
\\\cmidrule{2-3}
Actual & Non TC & TC  \\
\midrule
Non TC & TN & FP\\
 TC & FN &  TP\\
\hline
\end{tabular}
\label{truth}
\end{center}
\end{table}

where TN, TP, FN, FP are shown in Table~\ref{truth} for our classification problem.  A higher precision and recall are desirable. A higher precision indicates that a warning from the model for a possible TC formation can not be ignored, whereas a higher recall indicates that the model can detect a possible TC formation with a high probability. F1 score is a measure of the balance between precision and recall. 
To report the performance of the model in terms of these metrics, we have used 5-fold validation technique, whereas the dataset is partitioned into five equal subsets, and the model is evaluated on one subset after it is trained on the other four subsets. Finally, the average of five runs along with variation is reported for various leads time, which is called 5-fold validation accuracy. 

\section{Results and Analysis} \label{sec5}

\begin{table}[!h]
\begin{center}
\caption{5-fold performance ($\pm$std) of the model for T = 3 (12h).}
\begin{tabular}{p{0.8cm} p{0.9cm} p{1.7cm} p{1.7cm} p{1.7cm} p{1.7cm}}
\toprule
Ocean Basin & Lead Time(h) & Accuracy &  Precision & Recall & F1 \\
\midrule
NA & 24 & 0.943 $\pm$0.01 & 0.922 $\pm$0.02 & 0.965 $\pm$0.01 & 0.943 $\pm$0.01\\
 & 36 & 0.982 $\pm$0.01 & 0.981 $\pm$0.02 & 0.983 $\pm$0.01 & 0.982 $\pm$0.01\\
 & 48 & 0.977 $\pm$0.00 & 0.977 $\pm$0.01 & 0.976 $\pm$0.01 & 0.976 $\pm$0.00\\
 & 60 & 0.912 $\pm$0.02 & 0.925 $\pm$0.02& 0.901 $\pm$0.03 & 0.912 $\pm$0.02 \\
 \midrule
NI & 24 & 0.977 $\pm$0.01 & 0.964 $\pm$0.02 & 0.992 $\pm$0.01 & 0.978 $\pm$0.01\\
 & 36 & 0.990 $\pm$0.01 & 0.983 $\pm$0.01 & 0.997 $\pm$0.01 & 0.990 $\pm$0.01\\
 & 48 & 0.989 $\pm$0.01 & 0.984 $\pm$0.02 & 0.994 $\pm$0.01 & 0.989 $\pm$0.01\\
 & 60 & 0.929 $\pm$0.02 & 0.939 $\pm$0.03 & 0.918$\pm$0.03 & 0.928$\pm$0.02 \\
 \midrule
SI & 24 & 0.955 $\pm$0.01 & 0.932 $\pm$0.02 & 0.982 $\pm$0.01 & 0.956 $\pm$0.01\\
 & 36 & 0.993 $\pm$0.01 & 0.989 $\pm$0.01 & 0.996 $\pm$0.00 & 0.992 $\pm$0.00\\
 & 48 & 0.991 $\pm$0.01 & 0.995 $\pm$0.00 & 0.987 $\pm$0.01 & 0.991 $\pm$0.00\\
 & 60 & 0.913 $\pm$0.01 & 0.932 $\pm$0.02& 0.892 $\pm$0.02 & 0.912 $\pm$0.01 \\
\midrule
WP & 24 & 0.931 $\pm$0.01 & 0.920 $\pm$0.02 & 0.944 $\pm$0.02 & 0.931 $\pm$0.01\\
 & 36 & 0.975 $\pm$0.00 & 0.982 $\pm$0.01 & 0.968 $\pm$0.00 & 0.975 $\pm$0.00\\
 & 48 & 0.972 $\pm$0.01 & 0.978 $\pm$0.01 & 0.966 $\pm$0.01 & 0.972 $\pm$0.01\\
 & 60 & 0.869 $\pm$0.02 & 0.910 $\pm$0.03& 0.818 $\pm$0.02 & 0.862 $\pm$0.02 \\
\midrule

SP & 24 & 0.930 $\pm$0.03 & 0.898 $\pm$0.06 & 0.977 $\pm$0.03 & 0.933 $\pm$0.03\\
 & 36 & 0.964 $\pm$0.03 & 0.945 $\pm$0.05 & 0.992 $\pm$0.01 & 0.967 $\pm$0.02\\
 & 48 & 0.954 $\pm$0.04 & 0.926 $\pm$0.07 & 0.991 $\pm$0.01 & 0.956 $\pm$0.03\\
 & 60 & 0.899 $\pm$0.02 & 0.872 $\pm$0.05& 0.943 $\pm$0.05 & 0.903 $\pm$0.02 \\
 
\midrule
EP & 24 & 0.917 $\pm$0.01 & 0.884 $\pm$0.02 & 0.959 $\pm$0.01 & 0.920 $\pm$0.02\\
 & 36 & 0.966 $\pm$0.02 & 0.955 $\pm$0.02 & 0.979 $\pm$0.01 & 0.979 $\pm$0.01\\
 & 48 & 0.966 $\pm$0.01 & 0.963 $\pm$0.01 & 0.971 $\pm$0.02 & 0.967 $\pm$0.01\\
 & 60 & 0.904 $\pm$0.02 & 0.915 $\pm$0.02& 0.891 $\pm$0.02 & 0.903 $\pm$0.02 \\
 
 \bottomrule
\end{tabular}
\end{center}
\label{results1}
\end{table}

\begin{table}[!h]
\begin{center}
\caption{5-fold performance (std) of the model for T = 5 (24h).}
\begin{tabular}{p{0.8cm} p{0.9cm} p{1.7cm} p{1.7cm} p{1.7cm} p{1.7cm}}
\toprule
Ocean Basin & Lead Time(h) & Accuracy &  Precision & Recall & F1 \\
\midrule
NA & 24 & 0.953 $\pm$0.01 & 0.951 $\pm$0.02 & 0.956 $\pm$0.03 & 0.953 $\pm$0.01\\
 & 36 & 0.986 $\pm$0.01 & 0.993 $\pm$0.01 & 0.980 $\pm$0.02 & 0.987 $\pm$0.01\\
 & 48 & 0.913 $\pm$0.02 & 0.949 $\pm$0.02 & 0.873 $\pm$0.03 & 0.909 $\pm$0.02 \\
\midrule
NI & 24 & 0.974 $\pm$0.02 & 0.959 $\pm$0.03 & 0.992 $\pm$0.01 & 0.974 $\pm$0.02\\
 & 36 & 0.991 $\pm$0.01 & 0.988 $\pm$0.01 & 0.992 $\pm$0.01 & 0.990 $\pm$0.01\\
 & 48 & 0.925 $\pm$0.01 & 0.964 $\pm$0.03 & 0.941 $\pm$0.02 & 0.952 $\pm$0.01 \\
\midrule
SI & 24 & 0.960 $\pm$0.02 & 0.942 $\pm$0.03 & 0.982 $\pm$0.02 & 0.961 $\pm$0.01\\
 & 36 & 0.987 $\pm$0.01 & 0.986 $\pm$0.01 & 0.988 $\pm$0.01 & 0.987 $\pm$0.01\\
 & 48 & 0.936 $\pm$0.01 & 0.941 $\pm$0.03 & 0.933 $\pm$0.04 & 0.936 $\pm$0.01 \\

\midrule
WP & 24 & 0.935 $\pm$0.02 & 0.912 $\pm$0.04 & 0.965 $\pm$0.01 & 0.937 $\pm$0.02\\
 & 36 & 0.972 $\pm$0.02 & 0.966 $\pm$0.03 & 0.980 $\pm$0.01 & 0.973 $\pm$0.02\\
 & 48 & 0.888 $\pm$0.02 & 0.899 $\pm$0.04 & 0.877 $\pm$0.02 & 0.887 $\pm$0.02 \\
 
 \midrule
SP & 24 & 0.960 $\pm$0.01 & 0.936 $\pm$0.03 & 0.988 $\pm$0.01 & 0.961 $\pm$0.01\\
 & 36 & 0.980 $\pm$0.02 & 0.978 $\pm$0.02 & 0.984 $\pm$0.01 & 0.980 $\pm$0.02\\
 & 48 & 0.930 $\pm$0.01 & 0.940 $\pm$0.03 & 0.920 $\pm$0.04 & 0.928 $\pm$0.01 \\
 
  \midrule
EP & 24 & 0.949 $\pm$0.01 & 0.933 $\pm$0.02 & 0.968 $\pm$0.01 & 0.950 $\pm$0.01\\
 & 36 & 0.982 $\pm$0.01 & 0.980 $\pm$0.01 & 0.984 $\pm$0.01 & 0.982 $\pm$0.01\\
 & 48 & 0.919 $\pm$0.01 & 0.936 $\pm$0.01 & 0.900 $\pm$0.01 & 0.917 $\pm$0.01 \\
 \bottomrule
\end{tabular}
\end{center}
\label{results2}
\end{table}

The proposed model takes any 12h, or 24h (corresponding to T = 3, or 5) of continuous data at an interval of 6h as input from $t-72$ to $t-24$ hours, where $t$ is the time of possible TC formation, and predicts whether a TC will be formed or not.  This way for a particular $T$, the model predicts at lead times  $6k$, $4 \leq k \leq 13-T$. Thus the model predicts at a lead time of at least 24h, which is a minimum requirement for practical utility purpose.  For various leads time the model performance is reported in terms of the 5-fold accuracy along with the variation (std), in terms of above mentioned metrics in Tables~\ref{results1} and \ref{results2}. Increasing or decreasing values of $T$ do not improve the results further. 

The accuracy for 24h lead time, vary in a range of $91.7\% - 97.7\%$ and  $93.5\% - 97.4\%$ for $T$ equals to 3 and 5 respectively. So for a 24h lead time forecast, 36h (for T = 5) of data gives better results. The Precision remain quite high in the range of $88.4\%-96.4\%$ for both $T=3, 5$. This implies that model has a very small false alarm rate and any warning by model regarding possible TC formation can not be ignored. Also the Recall remains in the range of $94.4\% -99.2\%$ for both $T=3, 5$, which is quite high, implying that model is detecting nearly all TC formation cases, and it can be used reliably for practical purposes. The F1-score vary in the range of  $92.0\%-97.8\%$.

The accuracy for 36h lead time is even better than 24h lead time across all ocean basins, which is in the range $96.4\%-99.3\%$ and $97.2\%-99.1\%$ for $T$ equals to 3 and 5 respectively. A possible reason for this is that the reanalysis variables that we have selected in our study are more distinguishable and represent the TC formation well at this lead time. The Precision and Recall remain quite high again in the range of $94.5\%-99.3\%$ and $96.8\%-99.7\%$ respectively for both $T = 3, 5$.

The accuracy for lead time 48h is in the range $95.4\%-99.1\%$ which is greater than lead time 24h and slightly less than lead time 36h in the case of $T = 3$.  The accuracy for 48h lead time decreases by approx $5\%-9\%$ in comparison of lead time 24h and 36h for $T = 5$. This implies that for 48h lead time prediction $T = 3$ is a better choice. For $T=3$, model can predict with a lead time of 60h, which is quite a large time for early prediction of TC formation. In this case also model achieves an accuracy in the range of $86.9\%-92.9\%$, which can be considered good because the dynamics of causal factors behind TC formation change rapidly with time.  

\begin{figure}[!h]
    \centering
    \caption{Epoch wise train versus test accuracy for T = 3.}
    \includegraphics[width = 7cm, height = 5cm]{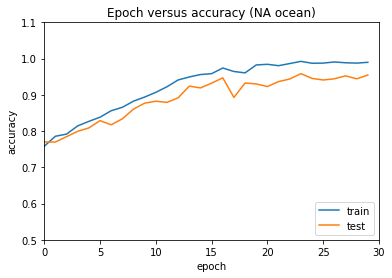}

    \label{epocNA}
\end{figure}

In Fig~\ref{epocNA} the epoch wise accuracy of the train and test set is shown for NA ocean basin. We achieved a similar train-test accuracy convergence in other ocean basins. 

\subsection{Comparison}

\begin{table}[!h]
\begin{center}
\caption{Comparision with best 5-fold performance reported in \cite{ptcgmc2019tzwy}.}
\begin{tabular}{p{1cm} p{1.5cm} p{1.8cm} p{1.8cm} p{1.8cm}}
\toprule
Ocean Basin & Lead Time(h)  &  Precision & Recall & F1 \\
\midrule
NA & 24 &  0.937 & 0.880 & 0.908 \\
 &  &  \textbf{0.951} & \textbf{0.965} & \textbf{0.953} \\
 & 48 &  0.888 & 0.683 & 0.757  \\
  &  &  \textbf{0.977} & \textbf{0.976} & \textbf{0.976} \\
\midrule
WP & 24 & 0.948  & 0.754  & 0.817 \\
 &  &  \textbf{0.920} & \textbf{0.965} & \textbf{0.937} \\
 & 48 & 0.889  &  0.642  & 0.701  \\
  &  &  \textbf{0.978} & \textbf{0.966} & \textbf{0.972} \\
 \bottomrule
\end{tabular}
\end{center}
\label{results3}
\end{table}

As discussed in section \ref{sec2}, the existing deep learning studies are not suitable for a direct comparison, as \cite{dldib2020sssgm,chen2019hybrid} deals with detecting an ongoing TC, and in \cite{matsuoka2018deep} the TC formation definition is based on cloud cover and uses simulated satellite data. We will make a direct comparison with the machine learning work \cite{ptcgmc2019tzwy}, where the TC formation definition coincides with our definition, and authors evaluated nine machine learning models in terms Precision, Recall, and F1-score for lead times up to 48h in NA, and WNP ocean basins. Out of nine classifiers, overall AdaBoost works best in all cases. In Table~\ref{results3}, we have reported the AdaBoost accuracy for lead times 24h and 48h along with accuracy achieved by our model (in bold).  From Table~\ref{results3}, we observe that for 24h lead time Precision is more or less same but there is a big difference in terms of Recall, whereas we achieve a recall value of $96.5\%$ both in NA and WP ocean basin in comparison of $88\%$ and $75.4\%$. In the case of lead time 48h, we achieve a better performance both in WP and NA ocean basins with an improvement of more than $9\%$ in precision and more than $28\%$ in the recall.  

\section{Conclusion} \label{sec6}

In this work a deep learning model is proposed which can forecast a TC formation using as less as 12h of data and lead time up to 60h with high precision, recall, and F1-score across six ocean basins of the world. An early information regarding potential cyclone formation has huge social, economical, and environmental benefits. Through this work, the authors establish that the reanalysis dataset has enough information to capture the complex and non-linear natural phenomenon behind a cyclone formation. One can further attempt to use the reanalysis dataset for a longer lead time forecast. The reanalysis dataset provides many other variables like cloud cover, vorticity, sea surface temperature, etc, one can explore these variables to further improve the model.  

One of the drawback of the proposed model is that we have taken equal number of positive and negative samples in our study. Given any random location and time, the chances of TC formation is very low. So a possible improvement of the proposed study is to take large negative samples than positive samples.












\bibliographystyle{unsrt}  

\bibliography{references}

\begin{thebibliography}{10}

\bibitem{pal1948for}
Erik Palmen.
\newblock On the formation and structure of tropical hurricanes.
\newblock {\em Geophysica}, 3(1):26--38, 1948.

\bibitem{elsberry1995tropical}
RUSSELL~L ELSBERRY.
\newblock Tropical cyclone motion. global perspectives on tropical cyclones.
\newblock {\em WMO/TD-No. 693}, pages 106--197, 1995.

\bibitem{gray1968global}
William~M Gray.
\newblock Global view of the origin of tropical disturbances and storms.
\newblock {\em Monthly Weather Review}, 96(10):669--700, 1968.

\bibitem{emanuel1989finite}
Kerry~A Emanuel.
\newblock The finite-amplitude nature of tropical cyclogenesis.
\newblock {\em Journal of Atmospheric Sciences}, 46(22):3431--3456, 1989.

\bibitem{tclkmm2019tcacc}
Thomas Knutson, Suzana~J. Camargo, Johnny C.~L. Chan, Kerry Emanuel, Chang-Hoi
  Ho, James Kossin, Mrutyunjay Mohapatra, Masaki Satoh, Masato Sugi, Kevin
  Walsh, and Liguang Wu.
\newblock Tropical cyclones and climate change assessment: Part i: Detection
  and attribution.
\newblock {\em Bulletin of the American Meteorological Society}, 100(10):1987
  -- 2007, 2019.

\bibitem{dvorak1975}
Vernon~F. Dvorak.
\newblock Tropical cyclone intensity analysis and forecasting from satellite
  imagery.
\newblock {\em Monthly Weather Review}, 103(5):420 -- 430, 1975.

\bibitem{dvorak1984}
Vernon~F Dvorak.
\newblock {\em Tropical cyclone intensity analysis using satellite data},
  volume~11.
\newblock US Department of Commerce, National Oceanic and Atmospheric
  Administration~…, 1984.

\bibitem{tcfg2013jc}
Joshua~H. Cossuth, Richard~D. Knabb, Daniel~P. Brown, and Robert~E. Hart.
\newblock Tropical cyclone formation guidance using pregenesis dvorak
  climatology. part i: Operational forecasting and predictive potential.
\newblock {\em Weather and Forecasting}, 28(1):100 -- 118, 2013.

\bibitem{tcgg2017myn}
Munehiko Yamaguchi and Naohisa Koide.
\newblock Tropical cyclone genesis guidance using the early stage dvorak
  analysis and global ensembles.
\newblock {\em Weather and Forecasting}, 32(6):2133 -- 2141, 2017.

\bibitem{halp2013eval}
Daniel~J Halperin, Henry~E Fuelberg, Robert~E Hart, Joshua~H Cossuth, Philip
  Sura, and Richard~J Pasch.
\newblock An evaluation of tropical cyclone genesis forecasts from global
  numerical models.
\newblock {\em Weather and Forecasting}, 28(6):1423--1445, 2013.

\bibitem{oeptc2009as}
Andrea~B. Schumacher, Mark DeMaria, and John~A. Knaff.
\newblock Objective estimation of the 24-h probability of tropical cyclone
  formation.
\newblock {\em Weather and Forecasting}, 24(2):456 -- 471, 2009.

\bibitem{henn2003for}
Christopher~C Hennon and Jay~S Hobgood.
\newblock Forecasting tropical cyclogenesis over the atlantic basin using
  large-scale data.
\newblock {\em Monthly weather review}, 131(12):2927--2940, 2003.

\bibitem{ddntd2015wzfm}
Wei Zhang, Bing Fu, Melinda~S. Peng, and Tim Li.
\newblock Discriminating developing versus nondeveloping tropical disturbances
  in the western north pacific through decision tree analysis.
\newblock {\em Weather and Forecasting}, 30(2):446 -- 454, 2015.

\bibitem{rs11101195}
Minsang Kim, Myung-Sook Park, Jungho Im, Seonyoung Park, and Myong-In Lee.
\newblock Machine learning approaches for detecting tropical cyclone formation
  using satellite data.
\newblock {\em Remote Sensing}, 11(10), 2019.

\bibitem{ptcgmc2019tzwy}
Tao Zhang, Wuyin Lin, Yanluan Lin, Minghua Zhang, Haiyang Yu, Kathy Cao, and
  Wei Xue.
\newblock Prediction of tropical cyclone genesis from mesoscale convective
  systems using machine learning.
\newblock {\em Weather and Forecasting}, 34(4):1035 -- 1049, 2019.

\bibitem{abjags2018phturnn}
Sheila Alemany, Jonathan Beltran, Adrian Perez, and Sam Ganzfried.
\newblock Predicting hurricane trajectories using a recurrent neural network.
\newblock {\em Proceedings of the AAAI Conference on Artificial Intelligence},
  33, 02 2018.

\bibitem{gspymjzsz2018anmpttlstm}
Song Gao, Peng Zhao, Bin Pan, Yaru Li, Min Zhou, Jiangling Xu, Shan Zhong, and
  Zhenwei Shi.
\newblock A nowcasting model for the prediction of typhoon tracks based on a
  long short term memory neural network.
\newblock {\em Acta Oceanologica Sinica}, 37:8--12, 05 2018.

\bibitem{kumar2021tptulstm}
Sandeep Kumar, Koushik Biswas, and Ashish~Kumar Pandey.
\newblock Track prediction of tropical cyclones using long short-term memory
  network.
\newblock In {\em 2021 IEEE 11th Annual Computing and Communication Workshop
  and Conference (CCWC)}, pages 0251--0257, 2021.

\bibitem{buobhlr2019}
Buo-Fu Chen, Boyo Chen, Hsuan-Tien Lin, and Russell~L. Elsberry.
\newblock Estimating tropical cyclone intensity by satellite imagery utilizing
  convolutional neural networks.
\newblock {\em Weather and Forecasting}, 34(2):447 -- 465, 2019.

\bibitem{mmrrmgieepti2020}
Manil Maskey, Rahul Ramachandran, Muthukumaran Ramasubramanian, Iksha Gurung,
  Brian Freitag, Aaron Kaulfus, Drew Bollinger, Daniel~J. Cecil, and Jeffrey
  Miller.
\newblock Deepti: Deep-learning-based tropical cyclone intensity estimation
  system.
\newblock {\em IEEE Journal of Selected Topics in Applied Earth Observations
  and Remote Sensing}, 13:4271--4281, 2020.

\bibitem{kumar2021track}
Sandeep Kumar, Koushik Biswas, and Ashish~Kumar Pandey.
\newblock Track prediction of tropical cyclones using long short-term memory
  network.
\newblock In {\em 2021 IEEE 11th Annual Computing and Communication Workshop
  and Conference (CCWC)}, pages 0251--0257. IEEE, 2021.

\bibitem{kumar2021imd}
Sandeep Kumar, Koushik Biswas, and Ashish~Kumar Pandey.
\newblock Prediction of landfall intensity, location, and time of a tropical
  cyclone.
\newblock In {\em Proceedings of the AAAI Conference on Artificial
  Intelligence}, volume~35, pages 14831--14839, 2021.

\bibitem{kumar2021reanalysis}
Sandeep Kumar, Koushik Biswas, and Ashish~Kumar Pandey.
\newblock Predicting landfall's location and time of a tropical cyclone using
  reanalysis data.
\newblock In {\em Artificial Neural Networks and Machine Learning -- ICANN
  2021}, pages 372--383. Springer International Publishing, 2021.

\bibitem{kumar2022will}
Sandeep Kumar, Koushik Biswas, and Ashish~Kumar Pandey.
\newblock Will a tropical cyclone make landfall?
\newblock {\em Neural Computing and Applications}, pages 1--12, 2022.

\bibitem{matsuoka2018deep}
Daisuke Matsuoka, Masuo Nakano, Daisuke Sugiyama, and Seiichi Uchida.
\newblock Deep learning approach for detecting tropical cyclones and their
  precursors in the simulation by a cloud-resolving global nonhydrostatic
  atmospheric model.
\newblock {\em Progress in Earth and Planetary Science}, 5(1):1--16, 2018.

\bibitem{sugi2002influence}
Masato Sugi, Akira Noda, and Nobuo Sato.
\newblock Influence of the global warming on tropical cyclone climatology: An
  experiment with the jma global model.
\newblock {\em Journal of the Meteorological Society of Japan. Ser. II},
  80(2):249--272, 2002.

\bibitem{yam2017res}
Yohei Yamada, Masaki Satoh, Masato Sugi, Chihiro Kodama, Akira~T Noda, Masuo
  Nakano, and Tomoe Nasuno.
\newblock Response of tropical cyclone activity and structure to global warming
  in a high-resolution global nonhydrostatic model.
\newblock {\em Journal of Climate}, 30(23):9703--9724, 2017.

\bibitem{dldib2020sssgm}
Snehlata Shakya, Sanjeev Kumar, and Mayank Goswami.
\newblock Deep learning algorithm for satellite imaging based cyclone
  detection.
\newblock {\em IEEE Journal of Selected Topics in Applied Earth Observations
  and Remote Sensing}, 13:827--839, 2020.

\bibitem{chen2019hybrid}
Rui Chen, Xiang Wang, Weimin Zhang, Xiaoyu Zhu, Aiping Li, and Chao Yang.
\newblock A hybrid cnn-lstm model for typhoon formation forecasting.
\newblock {\em Geoinformatica}, 23(3):375--396, 2019.

\bibitem{boussioux2020hurricane}
L{\'e}onard Boussioux, Cynthia Zeng, Th{\'e}o Gu{\'e}nais, and Dimitris
  Bertsimas.
\newblock Hurricane forecasting: A novel multimodal machine learning framework.
\newblock {\em arXiv preprint arXiv:2011.06125}, 2020.

\bibitem{gsygk2020tctufl}
Sophie Giffard-Roisin, Mo~Yang, Guillaume Charpiat, Christina Kumler~Bonfanti,
  Balázs Kégl, and Claire Monteleoni.
\newblock Tropical cyclone track forecasting using fused deep learning from
  aligned reanalysis data.
\newblock {\em Frontiers in Big Data}, 3:1, 2020.

\bibitem{era5cite}
Hans Hersbach, Bill Bell, Paul Berrisford, Shoji Hirahara, András Horányi,
  Joaquín Muñoz-Sabater, Julien Nicolas, Carole Peubey, Raluca Radu, Dinand
  Schepers, Adrian Simmons, Cornel Soci, Saleh Abdalla, Xavier Abellan,
  Gianpaolo Balsamo, Peter Bechtold, Gionata Biavati, Jean Bidlot, Massimo
  Bonavita, Giovanna De~Chiara, Per Dahlgren, Dick Dee, Michail Diamantakis,
  Rossana Dragani, Johannes Flemming, Richard Forbes, Manuel Fuentes, Alan
  Geer, Leo Haimberger, Sean Healy, Robin~J. Hogan, Elías Hólm, Marta
  Janisková, Sarah Keeley, Patrick Laloyaux, Philippe Lopez, Cristina Lupu,
  Gabor Radnoti, Patricia de~Rosnay, Iryna Rozum, Freja Vamborg, Sebastien
  Villaume, and Jean-Noël Thépaut.
\newblock The era5 global reanalysis.
\newblock {\em Quarterly Journal of the Royal Meteorological Society},
  146(730):1999--2049, 2020.

\bibitem{hodges2017well}
Kevin Hodges, Alison Cobb, and Pier~Luigi Vidale.
\newblock How well are tropical cyclones represented in reanalysis datasets?
\newblock {\em Journal of Climate}, 30(14):5243--5264, 2017.

\bibitem{ibtrackdata}
Kenneth~R. Knapp, Michael~C. Kruk, David~H. Levinson, Howard~J. Diamond, and
  Charles~J. Neumann.
\newblock The international best track archive for climate stewardship
  (ibtracs): Unifying tropical cyclone data.
\newblock {\em Bulletin of the American Meteorological Society}, 91(3):363 --
  376, 2010.

\bibitem{tsd2014mhk}
Michael Horn, Kevin Walsh, Ming Zhao, Suzana~J. Camargo, Enrico Scoccimarro,
  Hiroyuki Murakami, Hui Wang, Andrew Ballinger, Arun Kumar, Daniel~A.
  Shaevitz, Jeffrey~A. Jonas, and Kazuyoshi Oouchi.
\newblock Tracking scheme dependence of simulated tropical cyclone response to
  idealized climate simulations.
\newblock {\em Journal of Climate}, 27(24):9197 -- 9213, 2014.

\bibitem{6795724}
Y.~{LeCun}, B.~{Boser}, J.~S. {Denker}, D.~{Henderson}, R.~E. {Howard},
  W.~{Hubbard}, and L.~D. {Jackel}.
\newblock Backpropagation applied to handwritten zip code recognition.
\newblock {\em Neural Computation}, 1(4):541--551, 1989.

\bibitem{kasihg2012icdcnn}
Alex Krizhevsky, Ilya Sutskever, and Geoffrey Hinton.
\newblock Imagenet classification with deep convolutional neural networks.
\newblock {\em Neural Information Processing Systems}, 25, 01 2012.

\bibitem{7785132}
F.~{Milletari}, N.~{Navab}, and S.~{Ahmadi}.
\newblock V-net: Fully convolutional neural networks for volumetric medical
  image segmentation.
\newblock In {\em 2016 Fourth International Conference on 3D Vision (3DV)},
  pages 565--571, 2016.

\bibitem{lstm0}
Sepp Hochreiter and J\"{u}rgen Schmidhuber.
\newblock Long short-term memory.
\newblock {\em Neural Comput.}, 9(8):1735–1780, November 1997.

\bibitem{lstm1}
F.~A. {Gers}, J.~{Schmidhuber}, and F.~{Cummins}.
\newblock Learning to forget: continual prediction with lstm.
\newblock In {\em 1999 Ninth International Conference on Artificial Neural
  Networks ICANN 99. (Conf. Publ. No. 470)}, volume~2, pages 850--855 vol.2,
  1999.

\bibitem{lstm2}
Felix~A. Gers, Nicol~N. Schraudolph, and J\"{u}rgen Schmidhuber.
\newblock Learning precise timing with lstm recurrent networks.
\newblock {\em J. Mach. Learn. Res.}, 3(null):115–143, March 2003.

\bibitem{lstm3}
F.~A. {Gers} and E.~{Schmidhuber}.
\newblock Lstm recurrent networks learn simple context-free and
  context-sensitive languages.
\newblock {\em IEEE Transactions on Neural Networks}, 12(6):1333--1340, 2001.

\bibitem{chollet2015keras}
François Chollet.
\newblock Keras.
\newblock \url{https://github.com/fchollet/keras}, 2015.

\bibitem{relu}
Vinod Nair and Geoffrey~E. Hinton.
\newblock Rectified linear units improve restricted boltzmann machines.
\newblock In Johannes F{\"{u}}rnkranz and Thorsten Joachims, editors, {\em
  Proceedings of the 27th International Conference on Machine Learning
  (ICML-10), June 21-24, 2010, Haifa, Israel}, pages 807--814. Omnipress, 2010.

\bibitem{hjmsf1995inf}
Jun Han and Claudio Moraga.
\newblock The influence of the sigmoid function parameters on the speed of
  backpropagation learning.
\newblock In Jos{\'e} Mira and Francisco Sandoval, editors, {\em From Natural
  to Artificial Neural Computation}, pages 195--201, Berlin, Heidelberg, 1995.
  Springer Berlin Heidelberg.

\bibitem{scikit-learn}
F.~Pedregosa, G.~Varoquaux, A.~Gramfort, V.~Michel, B.~Thirion, O.~Grisel,
  M.~Blondel, P.~Prettenhofer, R.~Weiss, V.~Dubourg, J.~Vanderplas, A.~Passos,
  D.~Cournapeau, M.~Brucher, M.~Perrot, and E.~Duchesnay.
\newblock Scikit-learn: Machine learning in {P}ython.
\newblock {\em Journal of Machine Learning Research}, 12:2825--2830, 2011.

\bibitem{kdj2014adam}
Diederik Kingma and Jimmy Ba.
\newblock Adam: A method for stochastic optimization.
\newblock {\em International Conference on Learning Representations}, 12 2014.

\bibitem{tensorflow2015-whitepaper}
Mart\'{\i}n Abadi, Ashish Agarwal, Paul Barham, Eugene Brevdo, Zhifeng Chen,
  Craig Citro, Greg~S. Corrado, Andy Davis, Jeffrey Dean, Matthieu Devin,
  Sanjay Ghemawat, Ian Goodfellow, Andrew Harp, Geoffrey Irving, Michael Isard,
  Yangqing Jia, Rafal Jozefowicz, Lukasz Kaiser, Manjunath Kudlur, Josh
  Levenberg, Dandelion Man\'{e}, Rajat Monga, Sherry Moore, Derek Murray, Chris
  Olah, Mike Schuster, Jonathon Shlens, Benoit Steiner, Ilya Sutskever, Kunal
  Talwar, Paul Tucker, Vincent Vanhoucke, Vijay Vasudevan, Fernanda Vi\'{e}gas,
  Oriol Vinyals, Pete Warden, Martin Wattenberg, Martin Wicke, Yuan Yu, and
  Xiaoqiang Zheng.
\newblock {TensorFlow}: Large-scale machine learning on heterogeneous systems,
  2015.
\newblock Software available from tensorflow.org.

\end{thebibliography}

\end{document}